\documentclass[prd]{revtex4}

\usepackage{hyperref}
\usepackage{booktabs}

\usepackage{graphicx}
\usepackage{subfigure}

\usepackage{amsmath}
\usepackage{slashed}
\usepackage{SIunits}

\newcommand{\tr}[1]{\operatorname{tr} \left\{ #1 \right\}}

\begin{document}

\begin{abstract}
Within the framework of two-flavor covariant baryon chiral perturbation theory we have expressed the Chernyak-Zhitnitsky, Ioffe and Dosch currents in terms of chiral fields to provide leading one-loop extrapolation formulae for the leading and next-to-leading twist normalization constants $f_N$, $\lambda_1$ and $\lambda_2$. Finite volume effects due to pion loops have been taken into account. The occurring low energy constants are fitted to data obtained from recent lattice QCD simulations in order to extract the values at the physical point. 
\end{abstract}

\title{Chiral extrapolation of nucleon wave function normalization constants}
\author{Philipp \surname{Wein}}
\author{Peter C. \surname{Bruns}}
\author{Thomas R. \surname{Hemmert}}
\author{Andreas \surname{Sch\"afer}}
\affiliation{Universit\"at Regensburg}
\date{\today}
\maketitle

\section{Introduction}
In recent years, the combination of lattice quantum chromodynamics ((L)QCD) and effective field theories (EFTs) has proven to be especially fruitful. EFTs allow to relate hadronic properties in the real world to those in a fictitious one, where the quarks are heavier than in reality and are restricted to a finite volume (at least for not too small volumes). Hadron properties in such a theory are much cheaper to calculate numerically and this is what is usually done. This work addresses one specific item in this general broad program, namely the quark mass and volume dependence of the zeroth moments of specific distribution amplitudes (DAs), which are needed for the description of exclusive processes.\newline
The nucleon DAs are non-perturbative functions describing the internal nucleon structure. They are most conveniently defined by nucleon-to-vacuum matrix elements of nonlocal light-ray three-quark operators~\cite{braun:200642,Braun:2000kw,Chernyak:1984}. Restricting ourselves to contributions of (next-to-)leading twist and leading conformal spin, only local operators without derivatives have to be taken into account. Therefore only three independent structures remain, involving three (scale-dependent) non-perturbative constants $f_N$, $\lambda_1$ and $\lambda_2$, called leading and next-to-leading twist wave function normalization constants. \newline
Information on nucleon DAs can not be obtained in a direct way from experiments. They enter observables only through convolution integrals and theoretical guidance is needed to come up with sensible parametrizations. There are two theoretical frameworks capable of extracting information on DAs, which are of completely non-perturbative nature: the analytic QCD sum rule approach and the numeric approach of LQCD (see e.g.\ Ref.~\cite{Braun:QCDSF:2009}). The QCD sum rule technique is well established and leads to deep qualitative insights. However, it suffers from big systematic uncertainties, typically of order $10 \%-30 \%$. LQCD can provide high precision results, but only if all systematic uncertainties are well under control, including the quark mass dependence and finite volume effects.\newline %
The goal of this work is to set the chiral extrapolation of the wave function normalization constants mentioned above on a firm basis by the use of covariant baryon chiral perturbation theory (BChPT). The construction of the nucleon interpolating currents that project onto the normalization constants in terms of chiral fields forms the basis of this analysis. The currents corresponding to the next-to-leading twist normalization constants $\lambda_1$ and $\lambda_2$ appear also in grand unified theories, where they mediate proton decay, and have been constructed for the three-flavor case at leading order by Claudson, Hall and Wise in Ref.~\cite{Claudson:1982297}, which served us as starting point. Using the same technique we construct the three interpolating currents corresponding to the wave function normalization constants in a two-flavor framework and calculate the respective matrix elements to leading one-loop order, where we take into account finite volume effects due to pion loops. \newline
This work is organized as follows: in Section~\ref{sec_setup} we give a brief introduction to chiral perturbation theory (ChPT) and to nucleon interpolating currents. The construction of the currents in terms of chiral fields is performed in Section~\ref{sec_construction}. In Section~\ref{sec_results} we evaluate the contributing loop diagrams to obtain extrapolation formulae for the wave function normalization constants and fit the occurring low energy constants (LECs) to recent QCDSF lattice data in order to determine the values for physical masses. We conclude and give an outlook in  Section~\ref{sec_conclusion_and_outlook}.
\section{Theoretical Setup}\label{sec_setup}

\subsection{Covariant baryon chiral perturbation theory}
There exist quite a number of formulations of effective hadronic theories. In this work we use covariant BChPT for reasons of convenience: manifest Lorentz covariance simplifies the construction of the low energy form of nucleon interpolating currents and allows a direct comparison to the general decomposition of the proton-to-vacuum matrix elements given in Ref.~\cite{Braun:2000kw}.\newline
Chiral effective field theory (ChEFT) is based on the spontaneous breakdown of chiral symmetry in QCD. The additional (small) explicit symmetry breaking induced by finite quark masses yields massive quasi Goldstone bosons which can be interpreted as mesons. By establishing a connection between the set of left cosets of the unbroken vectorial subgroup in the chiral group ($G=SU(n_f)_L \otimes SU(n_f)_R$) and the Goldstone boson field configurations one finds that the mesonic degrees of freedom can be represented by a $SU(n_f)$ matrix $U$ which transforms linearly under a chiral rotation $g=(L,R) \in G$ (see Refs.~\cite{Leutwyler:1991mz,Scherer:2002tk}):
\begin{align}
U \rightarrow R U L^{\dagger} \ ,
\end{align}
i.e.\ the behavior of the Goldstone boson fields under symmetry transformations is determined by geometry (up to reparametrizations). It is most convenient and common practice to choose canonical coordinates for the parametrization of $U$~\footnote{The possibility of different parametrizations is unproblematic since all realizations of chiral symmetry are physically equivalent~\cite{Weinberg:1968de}.}. In the case of two light quark flavors ($n_f=2$) $U$ then reads:
\begin{align}
U \equiv e^{\frac{i}{F_{\pi}} \Phi} \quad \text{with} \quad \Phi \equiv \pi^i \sigma^i \ ,
\end{align}
where $F_{\pi}$ is the pion decay constant in the chiral limit, $\sigma^i \ (i=1,2,3)$ are the Pauli matrices and $\pi^i \ (i=1,2,3)$ represent the pion fields. The nucleon fields (proton and neutron) are grouped in a so-called isospinor
\begin{align}
\Psi \equiv \begin{pmatrix} p \\ n \end{pmatrix} \ .
\end{align}
Its transformation properties under the chiral group are only fixed on the vectorial subgroup where it has to transform as a doublet, i.e.\ we are free to choose the most convenient realization of chiral symmetry from various (physically equivalent \cite{Scherer:2002tk,Georgi:1985kw}) possibilities. Therefore we define the field
\begin{align}
u \equiv \sqrt{U} = e^{\frac{i}{2 F_{\pi}} \Phi} \ ,
\end{align}
which has to transform under chiral rotations as
\begin{align} \label{trafo_u}
u \rightarrow RuK^{\dagger} = KuL^{\dagger}
\end{align}
in order to reproduce the transformation properties of $U$. Eq.~\eqref{trafo_u} defines the compensator field $K$. Since $K$ meets the condition stated above, we can choose the nucleon field to transform as
\begin{align}
\Psi \rightarrow K \Psi \ .
\end{align}
This choice allows for a convenient construction of invariants because one can write down mesonic building blocks that transform as $X \rightarrow K X K^{\dagger}$, namely~\footnote{We neglect external sources since they are not of interest for our later evaluation of the overlap matrix elements.}:
\begin{align}
u_{\mu}&\equiv i \left( u^{\dagger} \partial_{\mu} u - u \partial_{\mu} u^{\dagger} \right) \ , \\
\chi_{\pm}&\equiv  u^{\dagger} \chi u^{\dagger} \pm u \chi^{\dagger} u \ ,
\end{align}
where $\chi = 2 B_0 \mathcal{M}$ includes the quark mass matrix $\mathcal{M}=\operatorname{diag}\left(m_u,m_d\right)$ and the condensate parameter $B_0$. The price one has to pay for adjusting the transformation properties of baryonic and mesonic fields this way is that non-vectorial chiral transformations are local since they depend on the mesonic field $u$. Hence one has to introduce covariant derivatives:
\begin{align}
D_{\mu} \Psi &\equiv \left( \partial_{\mu} + \Gamma_{\mu} \right) \Psi \ , \\
D_{\mu} X &\equiv  \partial_{\mu} X +  \left[ \Gamma_{\mu}, X \right] \ ,
\end{align}
where $\Gamma_{\mu}$ is called the chiral connection and is given as
\begin{align}
\Gamma_{\mu}\equiv\frac{1}{2}\left(u^{\dagger} \partial_{\mu} u + u \partial_{\mu} u^{\dagger} \right) \ .
\end{align}
In Table~\ref{building_blocks} we have summarized the building blocks together with their main properties for convenient reference. In the baryonic sector the nucleon three-momentum and the pion four-momentum (including the pion mass) serve as (small) parameters of the perturbative expansion generically denoted as $p$. The power of $p$ in a term defines its chiral dimension $D$. The lowest order pion-nucleon effective Lagrangian reads \cite{Gasser:1987rb}
\begin{align}
\mathcal{L}_{\pi N}^{(D=1)}=\bar{\Psi}\left(i\slashed{D}-m_0+\frac{g_A}{2} \slashed{u} \gamma_5 \right) \Psi \ ,
\end{align}
where $g_A$ is the axial coupling constant and $m_0$ is the nucleon mass (both in the chiral limit). For a full list of terms up to fourth order we refer the reader to Ref.~\cite{Fettes:2000gb}. To allow for both a systematic expansion of loop contributions in powers of $p$ and manifest Lorentz covariance we apply the infrared regularization (IR) scheme \cite{Becher:1999he}.
\begin{table}
\centering
\caption{\label{building_blocks}Chiral dimension and transformation properties of the basic fields. Note that the chiral dimension of a derivative depends on the field it acts on: it has dimension zero/one when acting on a nucleon/pion field.}
\renewcommand{\arraystretch}{1.5}
\begin{ruledtabular}
\begin{tabular}{l  c  c  c  c}
& $\Psi$	&  $u$	& $u_{\mu}$	& $\chi_{\pm}$\\ \hline
Chiral dimension		& 0		&	0	&	1		&	2\\
$SU(2)_L \otimes SU(2)_R$& $K\Psi$	&  $R u K^{\dagger} =  K u L^{\dagger}$	& $K u_{\mu} K^{\dagger}$& $K\chi_{\pm}K^{\dagger}$\\
Parity				& $\gamma_0 \Psi$& $u^{\dagger}$	& $(-1)^{\delta_{\mu0}}u_{\mu} $& $\pm\chi_{\pm}$\\
\end{tabular}
\end{ruledtabular}
\renewcommand{\arraystretch}{1.0}
\end{table}
\subsection{Three-quark interpolating currents}
We are interested in proton-to-vacuum matrix elements of three-quark currents. A general decomposition for such overlap matrix elements was given in Ref.~\cite{Braun:2000kw}. Restricting ourselves to contributions of (next-to-)leading twist and leading conformal spin, we can reduce the problem to the case where all quark fields are located at the origin. For this situation one obtains a simplified (local) decomposition of the matrix element:
\begin{align}\label{local_decomposition}
\begin{split}
\langle 0 | \epsilon^{ijk} u^i_{\alpha} u^j_{\beta} d^k_{\gamma} | N(k) \rangle &= \mathcal{V}_1^0 (\slashed{k}C)_{\alpha \beta} (\gamma_5 N)_{\gamma}+ \mathcal{V}_3^0 m_N (\gamma_{\mu}C)_{\alpha \beta}(\gamma^{\mu}\gamma_5 N)_{\gamma} \\
& \quad + \mathcal{T}_1^0 (k^{\nu} i \sigma_{\mu\nu}C)_{\alpha \beta}(\gamma^{\mu} \gamma_5 N)_{\gamma} + \mathcal{T}_3^0 m_N (\sigma_{\mu \nu}C)_{\alpha \beta} (\sigma^{\mu \nu} \gamma_5 N)_{\gamma} \ ,
\end{split}
\end{align}
where $| N(k) \rangle$ represents a proton state with four-momentum $k~(k^2=m_N^2)$, $N$ denotes the corresponding spinor and $u,d$ are quark field operators. The latin letters refer to color and the greek letters to Dirac indices. $C$ denotes the charge conjugation matrix. This decomposition has three independent parameters, since $\mathcal{V}_1^0=\mathcal{T}_1^0$ due to isospin constraints~\cite{Braun:2000kw}. In order to project onto the corresponding components we introduce the three standard baryonic currents: the Ioffe~(I), Dosch~(D) and Chernyak-Zhitnitsky~(CZ) current. They have been defined in~\cite{Chernyak:1984bm,Ioffe:1981kw,Chung:1981cc} as
\begin{subequations} \label{definition_of_currents}
\begin{align}
\eta^{\text{I}}(x)&\equiv\epsilon^{ijk}\left({u^{i}}^T(x)C\gamma_{\mu}u^j(x) \right) \gamma_5\gamma^{\mu}d^k(x) \ , \\
\eta^{\text{D}}(x)&\equiv\epsilon^{ijk}\left({u^{i}}^T(x)C\sigma_{\mu\nu}u^j(x) \right) \gamma_5\sigma^{\mu\nu}d^k(x) \ , \\
\eta^{\text{CZ}}(x)&\equiv\frac{2}{3}\epsilon^{ijk}\left[\left({u^{i}}^T(x)C\slashed{z}u^j(x) \right) \gamma_5\slashed{z}d^k(x) - \left({u^{i}}^T(x)C\slashed{z}d^j(x) \right) \gamma_5\slashed{z}u^k(x) \right] \ , \label{definition_CZ_current}
\end{align}
\end{subequations}
where $z$ is an arbitrary, dimensionless, light-like auxiliary vector. Note that the given definition of the Chernyak-Zhitnitsky current is not the original one but an isospin improved version which does not contain $\Delta^+$ baryon pollutions. The prefactor has been chosen such that the original current normalization is maintained. Using the decomposition~\eqref{local_decomposition} one obtains:
\begin{subequations} \label{current_matrix_elements}
\begin{align}
\langle 0 | \eta^{\text{I}}(0) | N(k) \rangle &=\left(\mathcal{V}_1^0-4\mathcal{V}_3^0 \right) m_N N \equiv \lambda_1 m_N N \ , \\
\langle 0 | \eta^{\text{D}}(0) | N(k) \rangle &=6\left(\mathcal{T}_1^0-4\mathcal{T}_3^0 \right) m_N N \equiv \lambda_2 m_N N \ , \\
\langle 0 | \eta^{\text{CZ}}(0) | N(k) \rangle &=\frac{1}{3}\left(\mathcal{V}_1^0+2\mathcal{T}_1^0\right) (z \cdot k) \slashed{z}N \equiv f_N (z \cdot k) \slashed{z}N \ .
\end{align}
\end{subequations}

\section{Construction of currents}\label{sec_construction}
In the following we will construct the currents given in Eq.~\eqref{definition_of_currents} in terms of the building blocks shown in Table~\ref{building_blocks}. For that purpose we analyze the symmetry properties of the currents under chiral rotations, parity transformation and (for the Chernyak-Zhitnitsky current) the separation in plus and minus components. At least for the two-flavor case the Ioffe and Dosch currents exhibit similar behavior under chiral transformations and thus can be treated within one approach. We refer to~\cite{Nagata:2007di} for a comprehensive presentation of the chiral transformation properties of three-quark operators.
\subsection{Ioffe and Dosch current}\label{sec_construction_ID}
Using the method of Fierz transformation~\cite{Fierz:1937,Nishi:2004st} one finds that the currents can be separated in parts of different helicity. Defining $q_p\equiv(1,0)^T$ to project onto the quark content of the proton we can rewrite the currents as
\begin{align}
\eta^{\text{I}}(x)&=4q^T_p \left( \gamma_R \mathcal{U}^L(x) - \gamma_L \mathcal{U}^R(x) \right) \ , \\
\eta^{\text{D}}(x)&=8q^T_p \left( \gamma_R \mathcal{U}^R(x) - \gamma_L \mathcal{U}^L(x) \right) \ ,
\intertext{where we have defined}
\mathcal{U}^{L/R}(x) &\equiv \epsilon^{ijk} \left( {q^i}^T_{L/R}(x)C(i\tau^2){q^j}_{L/R}(x) \right)q^k(x) \ .
\end{align}
$q\equiv(u,d)^T$ is a quark doublet field and $q_{L/R}\equiv \gamma_{L/R} \; q \equiv (1\mp \gamma_5)/2 \; q$ its left- and right-handed projection. $\tau^2$ is the second generator of the $SU(2)$ Lie algebra in the fundamental representation. By applying an infinitesimal chiral transformation to the operators $\gamma_L \mathcal{U}^{L/R}$ and $\gamma_R \mathcal{U}^{L/R}$ one finds that they transform as $(2_L,1_R)$ and $(1_L,2_R)$ respectively. Hence, we can write the low energy form of the currents as
\begin{align}
\eta^{\text{I}}&=4q_p^T \sum_{n=0}^{\infty} \sum_{k=1}^{i_n} \alpha_k^{(n)} \left( O^{(n)}_{k,RL} - O^{(n)}_{k,LR} \right) \ , \label{Ioffe_LE}\\
\eta^{\text{D}}&=8q_p^T \sum_{n=0}^{\infty} \sum_{k=1}^{i_n}  \beta_k^{(n)} \left( O^{(n)}_{k,RR} - O^{(n)}_{k,LL} \right) \ , \label{Dosch_LE}
\end{align}
where we have suppressed the $x$ dependence. $n$ gives the chiral dimension of the operator and $i_n$ is the number of operators with dimension $n$, which are labeled by $k$. Due to their similar transformation properties the Ioffe and Dosch currents only differ in the prefactor and the LECs $\alpha_k^{(n)},\beta_k^{(n)}$, i.e.\ $O^{(n)}_{k,RL}=O^{(n)}_{k,RR}$ and $O^{(n)}_{k,LR}=O^{(n)}_{k,LL}$. The different current operators that reproduce the transformation properties given above are listed in Table~\ref{LE_Ioffe_Dosch}, sorted by their chiral dimension. The different sign in some of the terms ensures correct (positive) parity. The terms which contribute to the proton-to-vacuum matrix element at leading one-loop order (which equals second chiral order) are marked by $(*)$.
\begin{table}[htb]
\centering
\caption{\label{LE_Ioffe_Dosch}Low energy operators for the Ioffe and Dosch current grouped according to their chiral dimension $n$. The list of second order terms is far from complete but solely contains terms that contribute to the proton-to-vacuum matrix elements of the currents at leading one-loop level (indicated by $(*)$). The contributions which vanish in the limit of exact isospin symmetry are marked with $(I)$.}
\renewcommand{\arraystretch}{1.5}
\begin{ruledtabular}
\begin{tabular}{c c  c  c  c }
$n$& $k$& $O^{(n)}_{k,RL}=O^{(n)}_{k,RR}$ & $O^{(n)}_{k,LR}=O^{(n)}_{k,LL}$ & \\ \hline
$0$&$1$ & $u \gamma_R \Psi$ & $-u^{\dagger} \gamma_L \Psi$ & $(*)$ \\
$1$&$1$ & $u u_{\mu}\gamma_R \gamma^{\mu} \Psi$ & $u^{\dagger} u_{\mu} \gamma_L \gamma^{\mu} \Psi$ &  \\
$1$&$2$ & $u u_{\mu}\gamma_R i D^{\mu} \Psi$ & $u^{\dagger} u_{\mu} \gamma_L i D^{\mu} \Psi$ &  \\
$2$&$1$ & $\tr{\chi_+}u \gamma_R \Psi$ & $-\tr{\chi_+}u^{\dagger} \gamma_L \Psi$ & $(*)$ \\
$2$&$2$ & $u \left( \chi_+ - \frac{1}{2} \tr{\chi_+} \right) \gamma_R \Psi$ & $-u^{\dagger} \left( \chi_+ - \frac{1}{2} \tr{\chi_+} \right) \gamma_L \Psi$ & $(*),(I)$ \\
& & $\vdots$ & $\vdots$ &
\end{tabular}
\end{ruledtabular}
\renewcommand{\arraystretch}{1.0}
\end{table}
\subsection{Chernyak-Zhitnitsky current}
For the construction of the current the transformation properties under chiral rotations play a key role. Splitting up the current into parts of different helicity by the use of Fierz transformation one obtains
\begin{align}
\eta^{\text{CZ}}(x)&=\frac{4}{3}q^T_p \tau^a \left( \gamma_L \tilde{\mathcal{U}}^{R,a}(x) - \gamma_R \tilde{\mathcal{U}}^{L,a}(x) \right) \ , \\
\intertext{where we have defined}
\tilde{\mathcal{U}}^{L/R,a}(x) &\equiv \epsilon^{ijk} z_{\mu} z_{\nu} \left( {q^i}^T_{L/R}(x)C\sigma^{\mu \rho}(i\tau^2)\tau^a{q^j}_{L/R}(x) \right) \sigma_{\rho}^{\ \nu}q^k(x) \ .
\end{align}
Applying an infinitesimal chiral transformation one finds that the diquark part does not transform in the trivial representation (as it was the case for the Ioffe and Dosch current) but under the three-dimensional adjoint representation. Therefore the operators $\gamma_L \tilde{\mathcal{U}}^{R,a}$ and $\gamma_R \tilde{\mathcal{U}}^{L,a}$ transform as $(2_L,3_R)$ and $(3_L,2_R)$ respectively. Suppressing the $x$ dependence the low energy version of the Chernyak-Zhitnitsky current takes the form
\begin{align}
\eta^{\text{CZ}}&=\frac{4}{3}q_p^T \tau^a \sum_{n=0}^{\infty} \sum_{k=1}^{i_n} \kappa_k^{(n)} \left( \tilde{O}^{(n),a}_{k,LR} - \tilde{O}^{(n),a}_{k,RL} \right) \ . \label{CZ_LE}
\end{align}
The operators $\tilde{O}^{(n),a}_{k,LR}$ and $\tilde{O}^{(n),a}_{k,RL}$ are listed in Table~\ref{LE_Chernyak_Zhitnitsky}. The relative sign is fixed by the condition that the total Chernyak-Zhitnitsky current has positive parity. The absence of terms with two derivatives acting on the nucleon field ($(z \cdot D)^2 \Psi$) needs some further comment. Consider the plus/minus projection operators which have been defined to separate components of lower and higher twist of a nucleon spinor with momentum $k$ (see e.g. Ref.~\cite{Braun:2000kw}):
\begin{align}
\Lambda_+ &\equiv \frac{\slashed{k}\slashed{z}}{2 z \cdot k } \ , &
\Lambda_- &\equiv \frac{\slashed{z}\slashed{k}}{2 z \cdot k } \ ,
\intertext{which, in position space, read}
\Lambda_+ &\equiv \frac{\slashed{\partial}\slashed{z}}{2 z \cdot \partial } \ , &
\Lambda_- &\equiv \frac{\slashed{z}\slashed{\partial}}{2 z \cdot \partial } \ .
\end{align}
Using the definition given in Eq.~\eqref{definition_CZ_current} it is easy to verify the following properties of the Chernyak-Zhitnitsky current:
\begin{align}
 \Lambda_+\eta^{\text{CZ}} &=0 \ , & \Lambda_-\eta^{\text{CZ}} &=\eta^{\text{CZ}} \ .
\end{align}
For an application of the plus/minus projectors on the low energy version of the current recall that the chiral connection $\Gamma_{\mu}$ is of first chiral order, since it contains a derivative acting on a meson field. Accordingly, we can replace the partial derivative by a covariant one (the difference is of higher order). Hence, we find that the minus projection of terms including two covariant derivatives acting on the nucleon has the form $\slashed{z}(z \cdot D)\Psi$:
\begin{align}
 \Lambda_-(z \cdot D)^2 \Psi = \frac{\slashed{z}\slashed{\partial}}{2 z \cdot \partial }(z \cdot D)^2 \Psi =\frac{\slashed{z}\slashed{D}}{2 z \cdot D }(z \cdot D)^2 \Psi + \mathcal{O}(p) \propto \slashed{z} (z \cdot D) \Psi + \mathcal{O}(p) \ ,
\end{align}
what justifies neglecting these terms in the operator. It is worth noting that for terms proportional to $\slashed{z}(z \cdot D)\Psi$ we have the exact relation
\begin{align}
 \Lambda_-\slashed{z} (z \cdot D) \Psi  = \frac{\slashed{z}\slashed{\partial}}{2 z \cdot \partial }\slashed{z} (z \cdot D) \Psi  =\left(1-\frac{\slashed{\partial}\slashed{z}}{2 z \cdot \partial }\right)\slashed{z} (z \cdot D) \Psi =\slashed{z} (z \cdot D) \Psi \ .
\end{align}
A similar argument leads to the structures $\slashed{z}z_{\mu} \Psi$ and $\slashed{z}(z \cdot D) \gamma_{\mu}\Psi$ for the terms of first chiral order.
\begin{table}[htb]
\centering
\caption{\label{LE_Chernyak_Zhitnitsky}Low energy operators for the Chernyak-Zhitnitsky current grouped according to their chiral dimension $n$. The list of second order terms is far from complete but solely contains terms that contribute to the proton-to-vacuum matrix element of the currents at leading one-loop level (indicated by $(*)$). The contributions which vanish in the limit of exact isospin symmetry are marked with $(I)$.}
\renewcommand{\arraystretch}{1.5}
\begin{ruledtabular}
\begin{tabular}{c c c  c  c }
$n$&$k$&  $\tilde{O}^{(n),a}_{k,LR}$ & $\tilde{O}^{(n),a}_{k,RL}$ & \\ \hline
$0$&$1$ & $z^{\mu}z^{\nu}{u^{\dagger}}^2 \tau^a u \gamma_L \gamma_{\mu} i D_{\nu} \Psi$ & $-z^{\mu}z^{\nu}u^2 \tau^a u^{\dagger} \gamma_R \gamma_{\mu} i D_{\nu}\Psi$ & $(*)$ \\
$1$&$1$  & $z^{\mu}z^{\nu}u^{\dagger} u_{\mu} u^{\dagger} \tau^a u \gamma_L \gamma_{\nu} \Psi$ & $z^{\mu}z^{\nu}u u_{\mu} u \tau^a u^{\dagger} \gamma_R \gamma_{\nu} \Psi$ &   \\
$1$&$2$ & $z^{\mu}z^{\nu}u^{\dagger} u^{\alpha} u^{\dagger} \tau^a u \gamma_L \gamma_{\mu} \gamma_{\alpha} i D_{\nu} \Psi$ &  $z^{\mu}z^{\nu}u u^{\alpha} u \tau^a u^{\dagger} \gamma_R \gamma_{\mu} \gamma_{\alpha} i D_{\nu} \Psi$ & \\
$1$&$3$ & $z^{\mu}z^{\nu}{u^{\dagger}}^2 \tau^a u u_{\mu}  \gamma_L \gamma_{\nu} \Psi$ & $z^{\mu}z^{\nu}u^2 \tau^a u^{\dagger} u_{\mu}  \gamma_R \gamma_{\nu} \Psi$ &   \\
$1$&$4$ &  $z^{\mu}z^{\nu}{u^{\dagger}}^2 \tau^a u u^{\alpha}  \gamma_L \gamma_{\mu} \gamma_{\alpha} i D_{\nu} \Psi$ & $z^{\mu}z^{\nu}u^2 \tau^a u^{\dagger} u^{\alpha}  \gamma_R \gamma_{\mu} \gamma_{\alpha} i D_{\nu} \Psi$ &  \\
$2$&$1$ & $z^{\mu}z^{\nu}\tr{\chi_+}{u^{\dagger}}^2 \tau^a u \gamma_L \gamma_{\mu} i D_{\nu} \Psi$ & $-z^{\mu}z^{\nu}\tr{\chi_+}u^2 \tau^a u^{\dagger} \gamma_R \gamma_{\mu} i D_{\nu}\Psi$ &  $(*)$ \\
$2$&$2$ &  $z^{\mu}z^{\nu}u^{\dagger} \left(\chi_+-\frac{1}{2} \tr{\chi_+}\right) u^{\dagger} \tau^a u \gamma_L \gamma_{\mu} i D_{\nu} \Psi$ &$-z^{\mu}z^{\nu}u \left(\chi_+-\frac{1}{2} \tr{\chi_+}\right) u \tau^a u^{\dagger} \gamma_R \gamma_{\mu} i D_{\nu}\Psi$ & $(*),(I)$ \\
$2$&$3$ & $z^{\mu}z^{\nu}{u^{\dagger}}^2 \tau^a u \left(\chi_+-\frac{1}{2} \tr{\chi_+}\right) \gamma_L \gamma_{\mu} i D_{\nu} \Psi$ & $-z^{\mu}z^{\nu}u^2 \tau^a u^{\dagger} \left(\chi_+-\frac{1}{2} \tr{\chi_+}\right) \gamma_R \gamma_{\mu} i D_{\nu}\Psi$ & $(*),(I)$ \\
& & $\vdots$ & $\vdots$ &
\end{tabular}
\end{ruledtabular}
\renewcommand{\arraystretch}{1.0}
\end{table}
\begin{figure}[htb]
\subfigure[\label{fg:subfig1}]{
\includegraphics[scale=0.55]{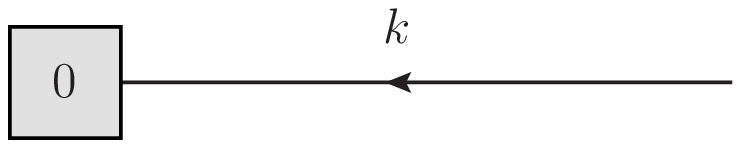}
}
\subfigure[\label{fg:subfig2}]{
\includegraphics[scale=0.55]{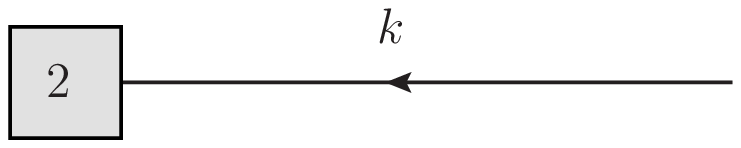}
}
\subfigure[\label{fg:subfig3}]{
\includegraphics[scale=0.55]{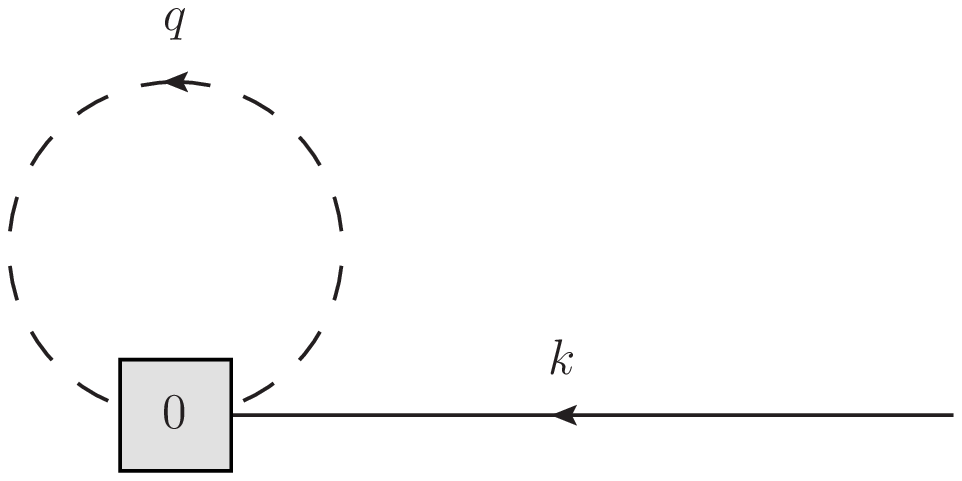}
}
\subfigure[\label{fg:subfig4}]{
\includegraphics[scale=0.55]{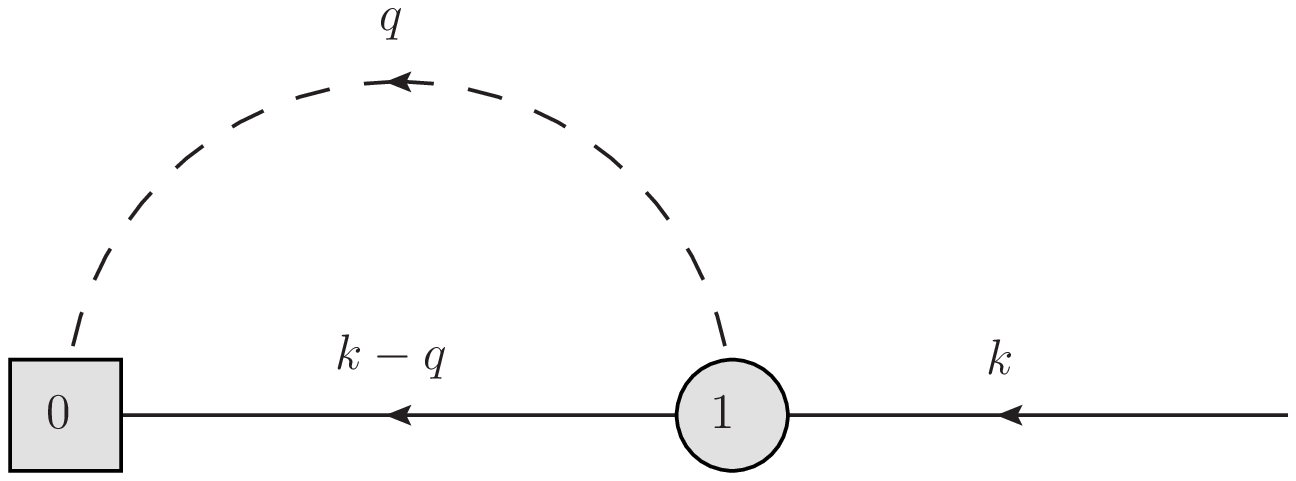}
}
\caption[]{\label{feynmangraphs}Feynman graphs relevant for the calculation of current proton-to-vacuum matrix elements to leading one-loop order. While \subref{fg:subfig1} is a diagram of zeroth chiral order, \subref{fg:subfig2}, \subref{fg:subfig3} and \subref{fg:subfig4} are graphs of second order.}
\end{figure}
\section{Results}\label{sec_results}
\subsection{Extrapolation formulae} \label{sec_Extrapolation_formulae}
Dimensional counting yields that one has to consider the Feynman diagrams shown in Fig.~\ref{feynmangraphs} to calculate the matrix elements given in Eq.~\eqref{current_matrix_elements} at leading one-loop level. The squares depicting the operators have to be derived from Eqs.~\eqref{Ioffe_LE}, \eqref{Dosch_LE} and \eqref{CZ_LE} for the respective current, while the circles denote standard vertices originating from the effective Lagrangian. The digits within the vertices and operators give their chiral dimension. The leading tree-level diagrams come with a $Z$-factor contribution. Since we work in the limit of exact isospin symmetry, the terms marked with $(I)$ in Tables~\ref{LE_Ioffe_Dosch} and \ref{LE_Chernyak_Zhitnitsky} do not contribute. The results for particular diagrams can be found in App.~\ref{app_results_by_diagram}. Interpreting the contribution of Fig.~\ref{fg:subfig2} as a chiral counterterm, i.e.\ splitting the corresponding LEC into a finite (denoted by ${}^{(r)}$) and a divergent part in such a way that the divergence of the loop graphs is cancelled exactly, we find the following extrapolation formulae:
\begin{subequations}\label{fit_formulae}
\begin{align}
\left(\lambda_1 m_N \right)(m_{\pi}) &= \alpha^{(0)}_1 \left( 4 - \frac{m_{\pi}^2}{2(4 \pi F_{\pi})^2} \left(6 g_A^2 +(3+9 g_A^2) \ln{\frac{m_{\pi}^2}{\mu^2}} \right)  \right) + 16 \alpha^{(2),(r)}_1 m_{\pi}^2 + \mathcal{O}(m_{\pi}^3) \ , \label{fit_formula_lambda1} \\
\left(\lambda_2 m_N \right)(m_{\pi}) &= \beta^{(0)}_1 \left( 8 - \frac{m_{\pi}^2}{(4 \pi F_{\pi})^2} \left(6 g_A^2 +(3+9 g_A^2) \ln{\frac{m_{\pi}^2}{\mu^2}} \right)  \right) + 32 \beta^{(2),(r)}_1 m_{\pi}^2 + \mathcal{O}(m_{\pi}^3) \ , \label{fit_formula_lambda2}
\\
f_N (m_{\pi}) &= \kappa^{(0)}_1 \left(1 - \frac{ m_{\pi}^2}{8(4 \pi F_{\pi})^2} \left( 6 g_A^2 +(19+9 g_A^2) \ln{\frac{m_{\pi}^2}{\mu^2}} \right)  \right) + 4 \kappa^{(2),(r)}_1 m_{\pi}^2 + \mathcal{O}(m_{\pi}^3) \ . \label{fit_formula_fN}
\end{align}
\end{subequations}
As expected we find an identical pion mass dependence (up to LECs) for $\lambda_1$ and $\lambda_2$ which is not surprising in view of the similar low energy form of the Ioffe and Dosch current (cf. Eqs.~\eqref{Ioffe_LE} and~\eqref{Dosch_LE}). The extrapolation formula for $f_N$ has a slightly different form as a consequence of a different weighting of the various graphs due to the particular form of the current (see App.~\ref{app_results_by_diagram}). Note that the formulae given above are only valid for an infinite volume. For a consideration of finite volume effects it is justified to apply the EFT formalism introduced in Refs.~\cite{Gasser:1986vb,Gasser:1987ah,Hasenfratz:1989pk,Gasser:1987zq}, since they are predominantly caused by mesons traveling around the box (contributions of single quarks walking around the box are exponentially suppressed \cite{Luscher:1985dn}). Therefore we rewrite $X \in \{ f_N, \lambda_1 m_N, \lambda_2 m_N \}$ as follows
\begin{align}\label{finite_volume}
X(m_{\pi},L_i=\infty) &= X(m_{\pi},L_i) + \underbrace{X(m_{\pi},L_i=\infty) - X(m_{\pi},L_i)}_{\equiv \delta X(m_{\pi},L_i)} \ ,
\end{align}
where the part with infinite boxlength is given by the corresponding formula of Eq.~\eqref{fit_formulae}, while $\delta X$ can be determined by the calculational method presented in Ref.~\cite{Hasenfratz:1989pk} (see App.~\ref{app_finite_volume_corrections} for the result). Note that by applying this procedure we consider the effect of pion loops, but neglect a volume dependence of the effective Lagrangian itself that cannot be excluded in our case: in Ref.~\cite{Gasser:1987zq} it was reasoned that for standard boundary conditions (i.e.\ periodic for bosonic and antiperiodic for fermionic fields in space and time direction) the finite and infinite volume effective Lagrangian densities resemble each other. However, this condition is not given for the lattice simulation we rely on. Like the majority of contemporary LQCD calculations it uses mixed boundary conditions for quark fields (antiperiodic in time and periodic in spatial directions). Therefore, strictly speaking, one cannot rely here on the proof given in Ref.~\cite{Gasser:1987zq} and one cannot exclude that the effective Lagrangian has to be modified by volume-dependent terms.
\subsection{Fitting of lattice data}
\begin{figure}[p!]
\subfigure[\label{ex:fN}]{
\includegraphics[scale=0.3]{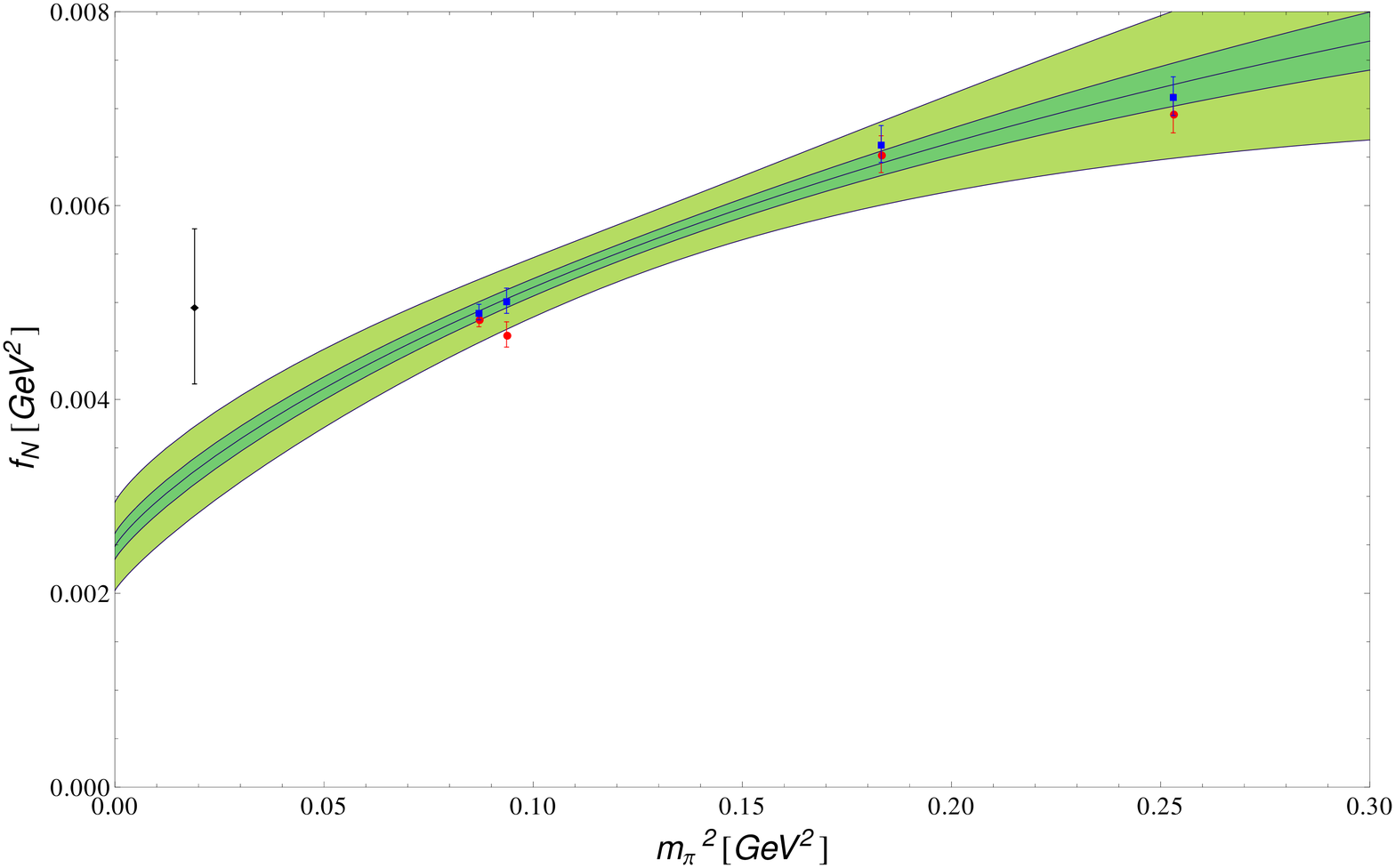}
}
\subfigure[\label{ex:l1mN}]{
\includegraphics[scale=0.3]{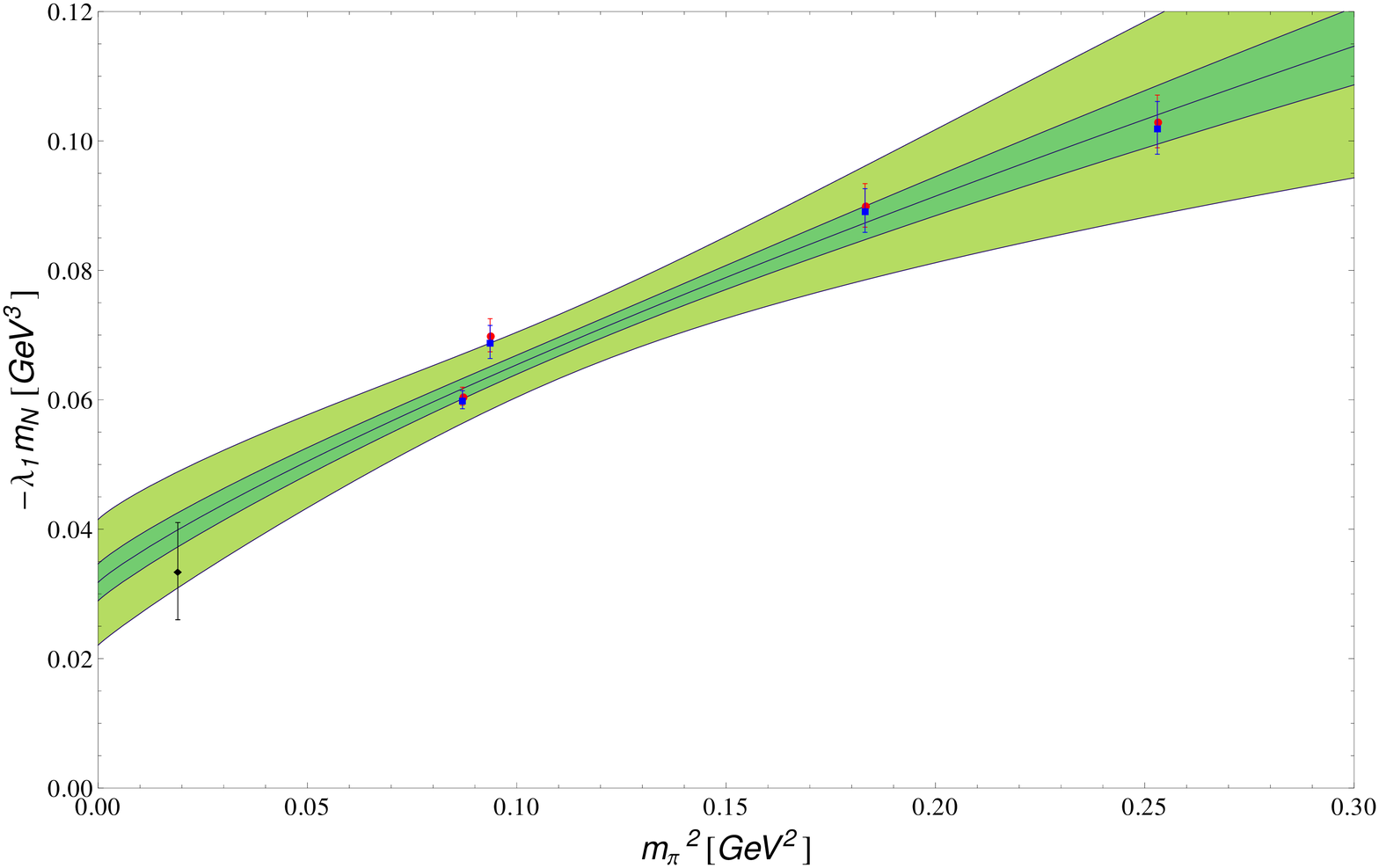}
}
\subfigure[\label{ex:l2mN}]{
\includegraphics[scale=0.3]{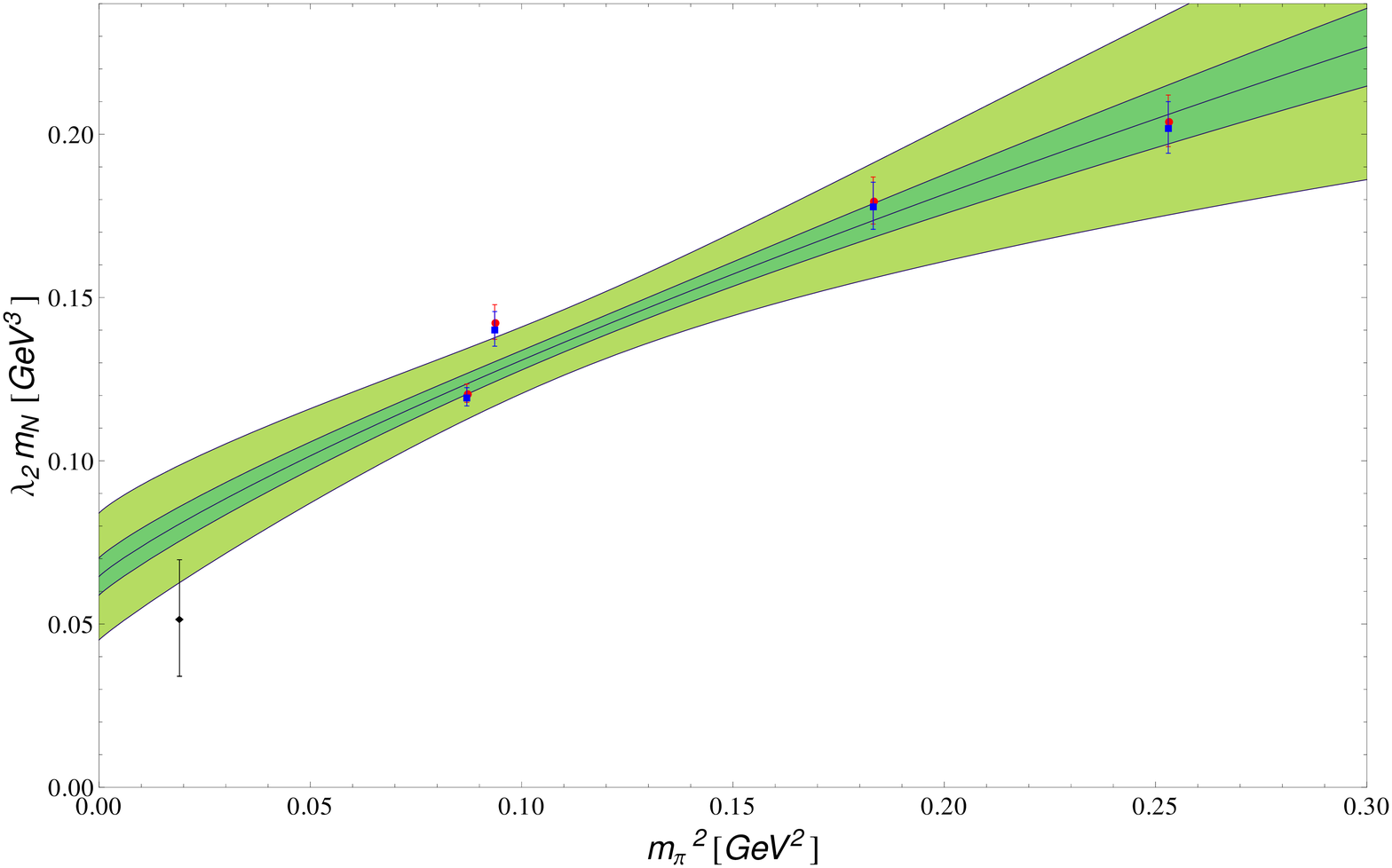}
}
\caption[]{\label{extrapolation}Fitted curves at infinite volume. The red points represent the original data points, while the blue ones are finite volume corrected. The black points at the physical pion mass are the sum rule predictions (NLO for \subref{ex:fN} and \subref{ex:l1mN}; LO for \subref{ex:l2mN}) and were not used for the fit. The colored bands represent the confidence bands for a confidence level of $68.3\%$ and $95.4\%$.}
\end{figure}

\begin{table}[tb]
\centering
\caption{\label{sum_rule_prediction}QCD sum rule predictions for $f_N$, $\lambda_1$ (NLO) and $\lambda_2$ (LO) at scales of $\unit{1}{\giga \electronvolt}$ (taken from Refs.~\cite{Braun:2000kw,Gruber:2010bj}) and $\unit{2}{\giga \electronvolt}$ (rescaled using a one-loop approximation).}
\renewcommand{\arraystretch}{1.5}
\begin{ruledtabular}
	\begin{tabular}{ c c c }
  &  $\unit{1}{\giga \electronvolt}$ &  $\unit{2}{\giga \electronvolt}$ \\ \hline
 $f_N \times 10^3 [\giga \electronvolt^{2}]$ & $5.1\pm0.8$ & $5.0\pm0.8$\\
 $-\lambda_1 \times 10^3 [\giga \electronvolt^{2}]\phantom{-}$ & $33\pm8\phantom{0}$ & $35\pm8\phantom{0}$ \\
 $\lambda_2 \times 10^3 [\giga \electronvolt^{2}]$ & $51\pm 19$ & $52\pm19$
	\end{tabular}
\end{ruledtabular}
\renewcommand{\arraystretch}{1.0}
\end{table}
In the following we will determine the occurring LECs by a fit to recent lattice data taken from Ref.~\cite{Schiel:2010hy,Schiel:priv2011}. Two flavors of clover fermions have been used for this lattice simulation with antiperiodic boundary conditions in the time direction and periodic boundary conditions in the spatial directions. The values are given at a scale of $\unit{2}{\giga \electronvolt}$ and the quarkonium parameter $r_0=\unit{0.5}{\femto\meter}$ has been used to determine the lattice spacing. For details on the lattice evaluation we refer to \cite{Braun:QCDSF:2009}. In future fits, using new yet unpublished data, QCDSF will use the derived ChPT expressions. The comparison presented here does not include these data and is meant as illustration for the general trend. Working in a ChEFT framework we have to restrict ourselves to pion masses that can be considered small with respect to a typical hadronic scale of $\unit{1}{\giga \electronvolt}$. For the given data set the particular value of $\unit{600}{\mega \electronvolt}$ constitutes a justifiable tradeoff between the accuracy of the fit and the reliability of (one-loop) ChPT (cf. Ref.~\cite{Djukanovic:2006xc}, where an upper limit of applicability of BChPT has been estimated to $\unit{600}{\mega \electronvolt}$). Table~\ref{sum_rule_prediction} contains the QCD sum rule values for $f_N$, $\lambda_1$ and $\lambda_2$ at scales of $\unit{1}{\giga \electronvolt}$ (taken from Refs.~\cite{Braun:2000kw,Gruber:2010bj}) and $\unit{2}{\giga \electronvolt}$ (rescaled in the one-loop approximation with $\Lambda_{\overline{\text{MS}}}=\unit{226}{\mega \electronvolt}$ following Ref.~\cite{Braun:QCDSF:2009} for better comparability). \newline
Since the extrapolation formulae~\eqref{finite_volume} describe the pion mass dependence of the product $\lambda_{1/2} m_N$ one might be tempted to divide them by $m_N$ and insert a chiral expansion for the nucleon mass (see for instance Ref.~\cite{Hemmert:2004}). However, a direct fit of the product $\lambda_{1/2}m_N$ turns out to be superior with respect to both effort and accuracy. The curves in Fig.~\ref{extrapolation} are obtained by fitting the LECs that appear in the extrapolation formulae~\eqref{finite_volume} to lattice data. Since the LECs only occur as pre\-factors, one can use a linear fit procedure. We have chosen the axial coupling $g_A=1.21$, the pion decay constant $F_{\pi}= \unit{86.2}{\mega \electronvolt}$ and the nucleon mass $m_0=\unit{0.88}{\giga \electronvolt}$ in the chiral limit~\cite{Bernard:2006te,Colangelo:2003hf,Hemmert:2004}. The fitted values for the relevant LECs are given in Table~\ref{fitted_LECs} while extrapolated values of the wave function normalization constants can be found in Table~\ref{nucleon_wave_function_normalization_constants}.
\begin{table}[tb]
\centering
\caption{\label{fitted_LECs} Fitted values for the LECs at leading one-loop order in the limit of exact isospin symmetry and $\mu=\unit{2}{\giga \electronvolt}$.}
\renewcommand{\arraystretch}{1.5}
\begin{ruledtabular}
	\begin{tabular}{ c c }
LEC & fitted value \\ \hline
$\kappa_1^{(0)} \times 10^3 [\giga \electronvolt^{2}]$ & $2.5\pm0.1$ \\
$\kappa_1^{(2)} \times 10^3 \phantom{[\giga \electronvolt^{0}]}$ & $0.6\pm0.5$ \\
$\alpha_1^{(0)} \times 10^3 [\giga \electronvolt^{3}]$ & $-7.9\pm0.6\phantom{-}$ \\
$\alpha_1^{(2)} \times 10^3 [\giga \electronvolt]$& $-10.2\pm1.8\phantom{-0}$ \\
$\beta_1^{(0)} \times 10^3 [\giga \electronvolt^{3}]$ & $8.0\pm0.6$ \\
$\beta_1^{(2)} \times 10^3 [\giga \electronvolt]$ & $9.8\pm1.8\phantom{}$ \\
 	\end{tabular}
\end{ruledtabular}
\renewcommand{\arraystretch}{1.0}
\end{table}
\begin{table}[tb]
\centering
\caption{\label{nucleon_wave_function_normalization_constants} Values for the wave function normalization constants at the physical point. Finite volume effects due to the pion cloud have been accounted for. The errors given are taken from the $95\%$ confidence bands. Note that we have neglected the systematic error introduced by the quarkonium parameter $r_0$.}
\renewcommand{\arraystretch}{1.5}
\begin{ruledtabular}
	\begin{tabular}{ c c }
normalization constant & value at the physical point \\ \hline
 $f_N \times 10^3 [\giga \electronvolt^{2}]$ & $3.3\pm0.5$ \\
 $-\lambda_1 \times 10^3 [\giga \electronvolt^{2}]\phantom{-}$& $43\pm10\phantom{}$\\
 $\lambda_2 \times 10^3 [\giga \electronvolt^{2}]$ & $86\pm19$
 	\end{tabular}
\end{ruledtabular}
\renewcommand{\arraystretch}{1.0}
\end{table}
\section{Conclusion and outlook}\label{sec_conclusion_and_outlook}
Within the framework of two-flavor covariant BChPT we have constructed the Ioffe, Dosch and Chernyak-Zhitnitsky current. While such a construction has already been performed by Claudson, Hall and Wise for the Ioffe and Dosch current in the three-flavor case~\cite{Claudson:1982297}, it is, to our knowledge, entirely new for the Chernyak-Zhitnitsky current. \newline
By calculating the zeroth moments of the respective nucleon DAs at leading one-loop order we provide chiral extrapolation formulae for the leading and next-to-leading twist wave function normalization constants $f_N$, $\lambda_1$ and $\lambda_2$. We took into account finite volume effects due to pion loops. Fitting the occurring LECs to recent lattice data we obtained the values at the physical point. The values obtained agree with QCD sum rule expectations except for $f_N$ which comes out somewhat small.\newline
To obtain information on the DAs at non-zero quark separation one has to consider higher moments of the DAs. These can be related to matrix elements of local operators containing derivatives acting on quark fields (see Ref.~\cite{Braun:QCDSF:2009}), which can be evaluated on the lattice. Therefore it is necessary to generalize our results to the case of operators containing derivatives in order to provide a systematic chiral extrapolation and finite volume corrections for the respective matrix elements. A possible application of the currents constructed in Sec.~\ref{sec_construction_ID} would be the evaluation of nucleon-to-pion transition DAs at leading one-loop level. Such a calculation seems promising because the result would depend on the same LECs (among others) and the matrix elements are relevant for the main proton decay channel $p \rightarrow \pi^0 + e^+$ in grand unified theories.
\begin{acknowledgments}
One of the authors wants to thank Rainer Schiel for the supply of up-to-date QCDSF lattice data and for helpful discussions. The QCDSF data used were generated on Hitachi SR8000 at LRZ (Munich), apeNEXT and APEmille at NIC/DESY (Zeuthen), BlueGene/Ls at NIC/JSC (J\"ulich), EPCC (Edinburgh) and KEK (by the Kanazawa group as part of the DIK research program) and QCDOC (Regensburg). This work was supported by DFG SFB/TR-55, by EU I3HP and by BMBF.
\end{acknowledgments}

\appendix
\section{Infrared integrals}
In order to apply the infrared regularization scheme (introduced in Ref.~\cite{Becher:1999he}) one defines the infrared integrals as follows:
\begin{align}
I_{k,l}&=\frac{\mu^{4-d}}{i} \int_{\text{IR}} \frac{d^dq}{(2 \pi)^d} \frac{1}{\left(m^2-q^2-i\epsilon \right)^k \left(M^2-(p-q)^2-i\epsilon \right)^l} \ ,
\end{align}
where $\mu$ denotes the scale and IR indicates that the Feynman parametrization connecting the meson and baryon propagators is extended to infinity. Using the definition
\begin{align}
L \equiv \frac{-1}{(4 \pi)^2} \left( \frac{1}{\epsilon}+\frac{1}{2} \left(1+\ln{4\pi}-\gamma_E \right) \right) \ ,
\end{align}
where $\epsilon\equiv 4-d$ and $\gamma_E$ is the Euler-Mascheroni constant, the solutions to $I_{1,0}$, $I_{0,1}$ and  $I_{1,1}$ read:
\begin{subequations}
\begin{align}
I_{1,0}&= 2 m^2 L + \frac{m^2}{(4 \pi)^2} \ln{\frac{m^2}{\mu^2}} \ , \\
I_{0,1}&= 0 \ , \\
I_{1,1}& = - \frac{m^2-M^2+p^2}{p^2} L + \frac{1}{(4 \pi)^2} \begin{aligned}[t] \Biggl[&\frac{m^2-M^2+p^2}{2p^2}\left(1-\ln{\frac{m^2}{\mu^2}} \right) \\ & - 2 \sqrt{\frac{m^2}{p^2}-\frac{\left(m^2-M^2+p^2 \right)^2}{4 p^4}} \arccos{\frac{-\left(m^2-M^2+p^2 \right)}{\sqrt{4 m^2 p^2}}} \Biggr] \ . \end{aligned}
\end{align}
\end{subequations}
For an evaluation of the tensorial integrals occurring in our calculations we define
\begin{subequations}
\begin{align}
p^{\mu} I_{k,l}^{(1)}&\equiv \frac{\mu^{4-d}}{i} \int_{\text{IR}} \frac{d^dq}{(2 \pi)^d} \frac{q^{\mu}}{\left(m^2-q^2-i\epsilon \right)^k \left(M^2-(p-q)^2-i\epsilon \right)^l} \ , \\
g^{\mu\nu} I_{k,l}^{(2)}+p^{\mu} p^{\nu} I_{k,l}^{(3)}&\equiv \frac{\mu^{4-d}}{i} \int_{\text{IR}} \frac{d^dq}{(2 \pi)^d} \frac{q^{\mu}q^{\nu}}{\left(m^2-q^2-i\epsilon \right)^k \left(M^2-(p-q)^2-i\epsilon \right)^l} \ .
\end{align}
\end{subequations}
Identities derived for standard loop integrals also hold for the infrared part, since the decomposition of standard loop integrals into an infrared singular part and a regular part is unique. Therefore Lorentz decomposition yields
\begin{subequations}
\begin{align}
I_{k,l}^{(1)}& = \frac{1}{2 p^2}\left( I_{k,l-1}-I_{k-1,l}+\left( m^2-M^2+p^2 \right) I_{k,l} \right) \ , \\
I_{k,l}^{(3)}& = \frac{1}{2 p^2(d-1)}\left( (2-d) I_{k-1,l}-2m^2 I_{k,l}+ d\left( m^2-M^2+p^2 \right) I_{k,l}^{(1)} \right) \ .
\end{align}
\end{subequations}
\section{Finite volume corrections}\label{app_finite_volume_corrections}
According to Ref.~\cite{Hasenfratz:1989pk} we obtain the finite volume correction $\delta X$ in Eq.~\eqref{finite_volume} by calculating the difference between integration and discrete sum over loop momenta (we neglect a possible volume dependence of the effective Lagrangian as mentioned in Sec.~\ref{sec_Extrapolation_formulae}). Hence, we define
\begin{align}
\delta I_{k,l}&= \frac{1}{i}\int\frac{dq_0}{2 \pi}\left\{\int\frac{d^3q}{(2 \pi)^3}-\frac{1}{\prod_i L_i} \sum_{\mathbf{q}} \right\}_{\text{IR}} \frac{1}{\left(m^2-q^2-i\epsilon \right)^k \left(M^2-(p-q)^2-i\epsilon \right)^l} \ ,
\end{align}
where $L_i$ is the boxlength in the $i$-th direction and the sum runs over the three-momenta $\mathbf{q}=2\pi\bigl(\frac{n_1}{L_1},\frac{n_2}{L_2},\frac{n_3}{L_3}\bigr)^T$ for integer $n_i$. Following Ref.~\cite{Hasenfratz:1989pk} we find
\begin{subequations}
\begin{align}
\delta I_{1,0}(L)& = \frac{-1}{(2 \pi)^2} \sum_{\mathbf{n}\neq 0 } \left( \frac{m^2}{\mathbf{l}_\mathbf{n}^2} \right)^{\frac{1}{2}} \operatorname{K}_{1}\left(\sqrt{ \mathbf{l}_{\mathbf{n}}^2 m^2} \right) \ , \\
\delta I_{0,1}(L)& = 0 \ , \\
\delta I_{1,1}(L)& = \frac{-1}{2 (2 \pi)^2} \int_0^{\infty} du \sum_{\mathbf{n} \neq 0 } \operatorname{K}_{0}\left(\sqrt{ \mathbf{l}_{\mathbf{n}}^2 f^2(u)} \right) e^{i u \mathbf{p} \cdot \mathbf{l}_{\mathbf{n}}} \ ,
\end{align}
\end{subequations}
where $\mathbf{l}_{\mathbf{n}}\equiv(n_1 L_1,n_2 L_2,n_3 L_3)^T$, $\operatorname{K}_i$ denote modified Bessel functions of the second kind and we have defined
\begin{align}
f^2(u) \equiv u M^2 + (1-u)m^2 + (u^2-u)p^2 \ .
\end{align}
For practical implementation we truncate the sum at $n_i=\pm4 \ (i=1,2,3)$.
\section{Results by diagram} \label{app_results_by_diagram}
In the following we present the results for the individual diagrams shown in Fig.~\ref{feynmangraphs} (at infinite volume). For the matrix element of the Ioffe current they read:
\begin{subequations}
 \begin{align}
\text{Amp}_a^{\text{I}} &= \sqrt{Z} 4 \alpha_1^{(0)} N(k) \ , \\
\text{Amp}_b^{\text{I}} &= 16 \alpha_1^{(2)} m_{\pi}^2 N(k) \ , \\
\text{Amp}_c^{\text{I}} &= -\frac{3\alpha_1^{(0)}}{2 F_{\pi}^2} I_{1,0} N(k) \ , \\
\text{Amp}_d^{\text{I}} &= \frac{3 \alpha_1^{(0)}g_A}{F_{\pi}^2} \left( I_{1,0} + (m_N^2-m_0^2) I_{1,1} - m_N (m_N+m_0) I_{1,1}^{(1)} \right) N(k) \ .
 \end{align}
\end{subequations}
One obtains the amplitudes for the Dosch current by the replacement $\alpha^{(n)}_k\rightarrow 2\beta^{(n)}_k$. The amplitudes of the Chernyak-Zhitnitsky current read:
\begin{subequations}
 \begin{align}
\text{Amp}_a^{\text{CZ}} &= \sqrt{Z} \kappa_1^{(0)} \slashed{z}(z \cdot k) N(k) \ , \\
\text{Amp}_b^{\text{CZ}} &= 4 \kappa_1^{(2)} m_{\pi}^2 \slashed{z}(z \cdot k) N(k) \ , \\
\text{Amp}_c^{\text{CZ}} &= -\frac{19 \kappa_1^{(0)}}{8 F_{\pi}^2}  I_{1,0} \slashed{z} (z \cdot k) N(k)\ , \\ 
\text{Amp}_d^{\text{CZ}} &=  \frac{5 \kappa_1^{(0)} g_A}{4 F_{\pi}^2} \left(  I_{1,0} + (m_N^2-m_0^2)\left( I_{1,1} - I_{1,1}^{(1)}\right)- m_N(m_N+m_0)\left( I_{1,1}^{(1)} - I_{1,1}^{(3)}\right) \right) \slashed{z}(z \cdot k)  N(k) \ .
 \end{align}
\end{subequations}
The $Z$-factor contribution has the form:
\begin{align}
\sqrt{Z} = 1 + \frac{3 g_A^2}{8 F_{\pi}^2}\frac{\partial}{\partial \slashed{p}} \left( m_{\pi}^2 \left( \slashed{p} + m_N \right) I_{1,1} + \slashed{p}\left( m_N^2-\slashed{p}^2 \right) I_{1,1}^{(1)} \right)\Bigr|_{\slashed{p} = m_N} + \mathcal{O}(m_{\pi}^3)\ .
\end{align}
\bibliographystyle{utphys}
\bibliography{bibliography}

\end{document}